Article

# Collaborating with communities: Citizen Science Flood Monitoring in Urban Informal Settlements

Erich Wolff [1,*], Matthew French [2], Noor Ilhamsyah [3], Mere Jane Sawailau [4] and Diego Ramírez-Lovering [1]

[1] Faculty of Art, Design and Architecture, Monash University, Melbourne, Australia; E-Mail: erich.wolff@monash.edu

[2] Monash Sustainable Development Institute, Monash University, Melbourne, Australia;

[3] Revitalising Informal Settlements and their Environments Program, Makassar, Indonesia;

[4] Revitalising Informal Settlements and their Environments Program, Suva, Fiji;

* Corresponding author

**Abstract**

Concerns regarding the impacts of climate change on marginalised communities in the Global South have led to calls for affected communities to be more active as agents in the process of planning for climate change. While the value of involving communities in risk management is increasingly accepted, the development of appropriate tools to support community engagement in flood risk management projects remains nascent. Using the Revitalising Informal Settlements and their Environment (RISE) Program as a case study, the article interrogates the potential of citizen science to include disadvantaged urban communities in project-level flood risk reduction planning processes. This project collected more than 5000 photos taken by 26 community members living in 13 informal settlements in Fiji and Indonesia between 2018 and 2020. The case study documents the method used as well as the results achieved within this 2-year project. It discusses the method developed and implemented, outlines the main results, and provides lessons learned for others embarking on citizen science environmental monitoring projects. The case study indicates that the engagement model and the technology used were key to the success of the flood-monitoring project. The experiences with the practice of monitoring floods in collaboration with communities in Fiji and Indonesia provide insights into how similar projects could advance more participatory risk management practices. The article identifies how this kind of approach can collect valuable flood data while also promoting opportunities for local communities to be heard in the arena of risk reduction and climate change adaptation.

**Keywords**
Citizen science; climate change, community-based methods; disasters; floods; informal settlements

## 1. Introduction

The notion that disaster studies must better account for the needs of disadvantaged communities has been gaining traction since the 2030 Agenda for Sustainable Development's call for "leaving no one behind" (UN, 2015). This is reflected in arguments for the involvement of communities in data collection to respond to important challenges in the sustainable development agenda (Fritz et al., 2019). Among growing concerns with floods and other environmental hazards, citizen science has emerged as a promising approach for involving communities in disaster risk management (Cooper et al., 2021). Investigating how community members can contribute to more inclusive risk management practices is particularly important in the context of informal settlements, which are expected to be disproportionately affected by the impacts of climate change (French et al., 2020; Hoegh-Guldberg et al., 2018).

Risk management is formally understood as the assessment, evaluation and intervention on the potential of a particular threat to cause results that differ from expected outcomes in a specific system (ISO/TC 262, 2018). Since the conceptualisation of risk, the practice of risk assessment has been a key stage in the process of managing uncertainty and mitigating the impacts of disasters (Renn 1992). Historically, at the turn of the 21st century, the work of thinkers such as Beck (1992) and Giddens (1999) shed light on how risk management had defined cities and shaped planning practices globally. These seminal works denounced, for the first time, the limitations of the "risk society" while calling for a more reflexive and inclusive practice in the management of uncertainty. Now, decades later, these discussions gain traction again as societies struggle to address the growing and increasingly uncertain risks in the wake of the climate crisis.

It has been argued that the policies and practices of risk mitigation have been primarily defined by a few risk experts whose recommendations play a disproportionate role in decision-making processes (Knowles, 2011). In flood-prone areas, for instance, traditional risk management frameworks suggest that land use should be guided by the assessment of water level fluctuations, which are quantified and assessed through the methods and language of risk (Olesen et al., 2017). Emerging perspectives claim that these frameworks limit the involvement of local communities and might not be easily applicable in understudied contexts (Kuhlicke et al., 2020). As such, the use of citizen-generated data enabled by the democratisation of the internet has gained traction within the field of disaster studies. These changes in the field suggest that approaches to risk management are being gradually transformed to better account for the challenges of informal settlements, for which flood data is often unavailable.

Characterised by insecure land tenure, lack of access to infrastructure and non-conformity with regulatory frameworks (UN Habitat III, 2017), informal settlements are also expected to be disproportionately impacted by climate change due to their rapid urban growth over areas at high risk from extreme weather (Bettini et al., 2017; Chandler, 2019; French et al., 2020; Revi et al., 2014). Some works have critiqued the social and political implications of the application of traditional risk management in informal settlements (French et al., 2020; Sandoval & Sarmiento, 2020; Yarina, 2018), which have historically been severely affected by floods and tropical cyclones. Emerging as an alternative, community-based approaches consider communities not as clients or external consultants, but as central agents in risk identification, monitoring and communication (Shaw, 2014).

In response to these conditions, there have been growing calls for more direct involvement of local communities in risk management (Kelman, 2019). A reconsideration of the practices through which risks have been conceptualised and managed is particularly necessary for the context of low- and middle-income countries. This has been recognised, for example, in the United Nations' Sendai Framework for Disaster Risk Reduction which acknowledges that special attention should be dedicated to providing resources and expertise for the management of disasters in the Global South, particularly in island nations (UNISDR, 2015).

In this context, this article explores an empirical case study of a citizen science project conducted in Indonesia and Fiji from the perspective of RISE community fieldworkers directly involved in the implementation of the project. Having a unique perspective of the project, the experiences of fieldworkers are critical to investigate how risk management practices can effectively involve disadvantaged urban communities in the process of making sense of disasters. In doing so, this article aims to provide insights into the operation and implementation aspects of community-based flood-monitoring practices by discussing the firsthand lessons from a citizen science project within the broader Revitalising Informal Settlements and their Environments (RISE) program.

## 2. The emergence of citizen science

The expertise to map, understand and monitor environmental hazards has been increasingly considered a strategic asset in a world characterised by growing risks. The application of risk management as a practice to address environmental hazards, such as landslides, floods and cyclones, has become ubiquitous in high-income countries, but comparatively less progress has been made in the provision of basic infrastructure to marginalised communities (*Habitat III Issue Paper*, 2015). For over two decades now, United Nations organisations have been calling for more resources to address the challenges of infrastructure provision and risk management in informal settlements (*Habitat III Issue Paper*, 2015; UN Habitat, 2003; UNDRR/CRED, 2020; UNISDR, 2015). Researchers have also been suggesting that the unequal access to the resources, political power and expertise to manage hazards affects the just distribution of opportunities and risks in the city (Anguelovski et al., 2016). In response to these concerns, the field of disaster studies has been increasingly interested in involving communities in the process of collecting environmental data to inform climate adaptation projects (See, 2019; Sy et al., 2020).

Researches studying the emergence of smart cities argue that the democratisation of communication technologies, such as mobile internet and smartphones, has created new possibilities for the use of citizen-generated data (Townsend, 2015). Among these approaches, the involvement of citizens in the process of gathering environmental data through citizen science projects is seen as particularly promising (See, 2019; Starkey et al., 2017; Wolff & Muñoz, 2021). The use of citizen science within the field of hydrology and flood risk management is based on involving non-scientists in the process of characterising and monitoring floods, commonly under the supervision of a scientific body or practitioner (Haklay, 2015). The growing popularity of citizen science can be attributed not only to the capacity of these methods to generate cheap, up-to-date and accessible disaster risk data but also to the perceived social engagement benefits of participatory approaches (Haklay et al., 2018).

While digital technologies have increasingly facilitated the collection of citizen-generated data (Karvonen, 2020), the literature shows that the participation of community members in the process of monitoring and assessing risks is not a recent phenomenon. In practice, the concepts of citizen science and community-based monitoring can be traced back to the notion of civic science (Kruger & Shanno, 2000) or to the early concept of people's science (Wisner et al., 1977), which encouraged communities to contribute to the documentation of a particular phenomenon of scientific interest. The group of approaches that seek to collect data in partnership with citizens include volunteered geographic information (Haworth et al., 2018), crowdsourced data (Lowry & Fienen, 2013), community science (Carr, 2004) and citizen science (Haklay, 2015).

Although citizen-generated datasets are considered promising within the flood risk assessment literature (Le Coz et al., 2016; See, 2019; Voinov & Gaddis, 2008), their use is still limited to simple applications. The literature shows that this kind of data has been used in the context of disaster management primarily for validating prediction models (Starkey et al., 2017). Other applications rely on citizen-generated data for conducting emergency assessments of the intensity and extent of occurrences for post-disaster planning (Fohringer et al., 2015; Smith & Rodriguez, 2017). As the notion of participatory science gains currency, however, it is important to consider that community-based approaches can play other roles and foster a more people-centred approach to disaster risk management.

Projects dedicated to mapping water level fluctuations with local-scale precision and continued engagement with community members are expected to become increasingly common. Despite a growing interest in this approach, the practical and operational aspects of employing citizen science to monitor floods in urban informal settlements are still largely undocumented to date. In an effort to address this deficit in knowledge, this article reflects on the operational aspects of RISE's flood-monitoring project in Fiji and Indonesia to document the methods and tools used and to provide lessons on how to implement citizen science in urban informal settlements.

## 3. Case Study: Flood-monitoring in Fiji and Indonesia

The Revitalising Informal Settlements and their Environments (RISE) program is a transdisciplinary research initiative implementing nature-based infrastructure systems in the Asia-Pacific (Brown et al., 2018; Ramirez-Lovering et al., 2018). Its primary research aims to investigate the human health and environmental effects of nature-based infrastructure in 12 settlements in Suva (Fiji) and 12 settlements in Makassar (Indonesia). Using a Randomised Control Trial model, RISE selected settlements that would allow researchers to study the benefits of the infrastructure systems in both countries. The selected sites represent a diversity of characteristics with varying biophysical, socioeconomic and land tenure conditions (Leder et al. 2021). Varying considerably in terms of size and physical settings, the population of the settlements ranges between 50 and 700 people each and include sites with coastal, riverine and flood plain characteristics.

Developed within the broader RISE program, the flood-monitoring project was designed to document floods in the most flood-prone sites: seven settlements in Suva and six settlements in Makassar. This project was key to informing the design of RISE's nature-based infrastructure since the wetland systems need to be protected from direct damage caused by flooding (ADB and RISE, 2021). Furthermore, it was critical to monitor floods in the sites because water level variations were identified as a potential source of contamination capable of affecting human health in the settlements (French et al., 2021). Considering the limited information available for modelling floods on the sites, RISE researchers developed this project to systematically monitor floods in the participating settlements.

The model used is similar to several other data-crowdsourcing and distributed intelligence initiatives that invited citizens in the collection and analysis of flood data over the last decade (Fava et al., 2018; Kankanamge et al., 2020; Le Coz et al., 2016; Mobley et al., 2019; Smith & Rodriguez, 2017). Other projects conducted in low- and middle-income countries influenced the data collection model used as they provided important insights into the challenges of collecting flood data in "data-poor" contexts such as informal settlements (Adomah Bempah & Olav Øyhus, 2017; Glas et al., 2020; Hazarika et al., 2018). In Indonesia, the PetaBencana project (*PetaBencana*, 2021) was an important precedent because it exemplifies how citizen-generated data can inform decision-making processes during and after floods (Fadmastuti, 2019).

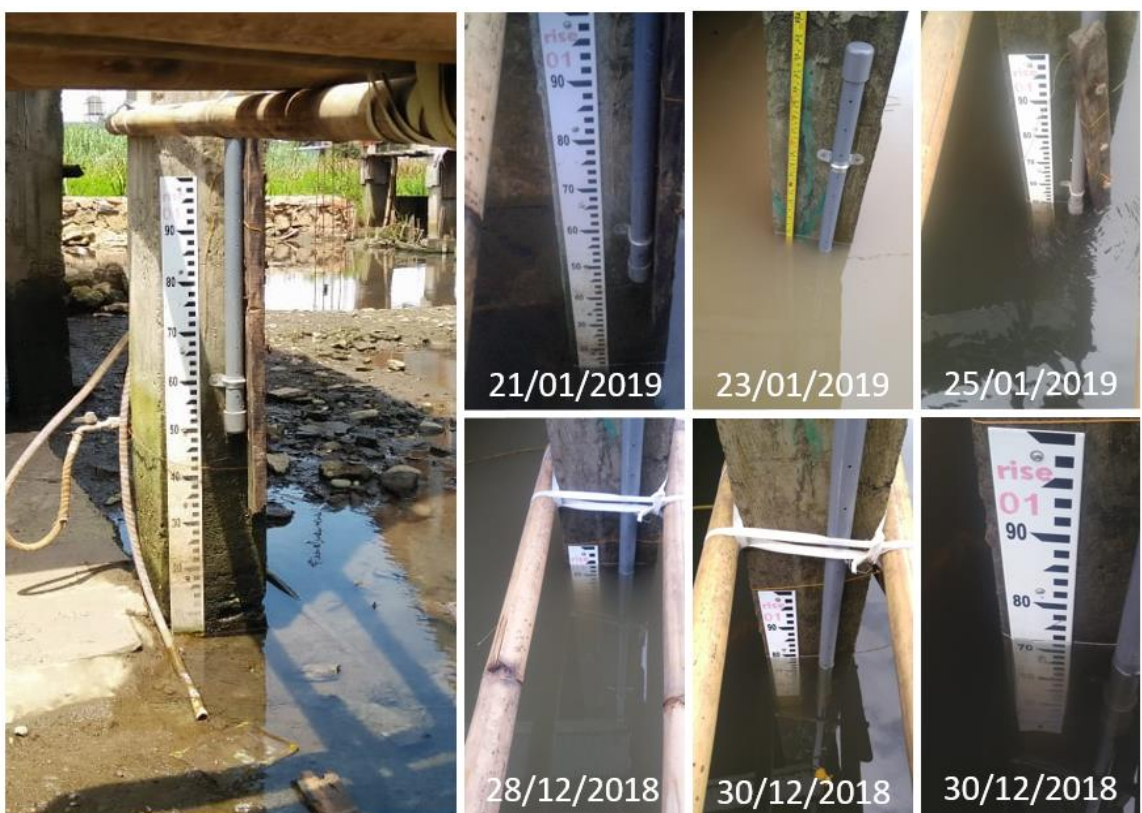

**Figure 1.** Flood gauge in Makassar, Indonesia, photographed by residents in different conditions during floods in 2018 and 2019.

Adapting the methods utilised in these projects to the context of RISE, the researchers conceptualised this initiative as a repository of flood photos that would later be interpreted to provide evidence of flood levels, as shown in Figure 1. As such, the project was conceptualised as a data crowdsourcing initiative in the sense that it involved residents in a role analogous to that of a "sensor" collecting flood data (Haklay, 2013). After its implementation, however, it became evident that citizens were playing other roles in the project as they supported the analysis of the photos and actively contributed to disseminating the results among the communities.

The preparation for the project consisted of the installation of a gauge and a crest level indicator in each of the flood-prone settlements in a position where the participants could safely photograph the gauges and register water level fluctuations. The participants were instructed to use their personal smartphones to send photos of the flood gauges daily (as shown in Figure 2) in order to keep a record of the water levels throughout the whole season. The RISE staff members that engaged the communities instructed the participants to photograph the gauges at least once a day and periodically at two-hour intervals during floods.

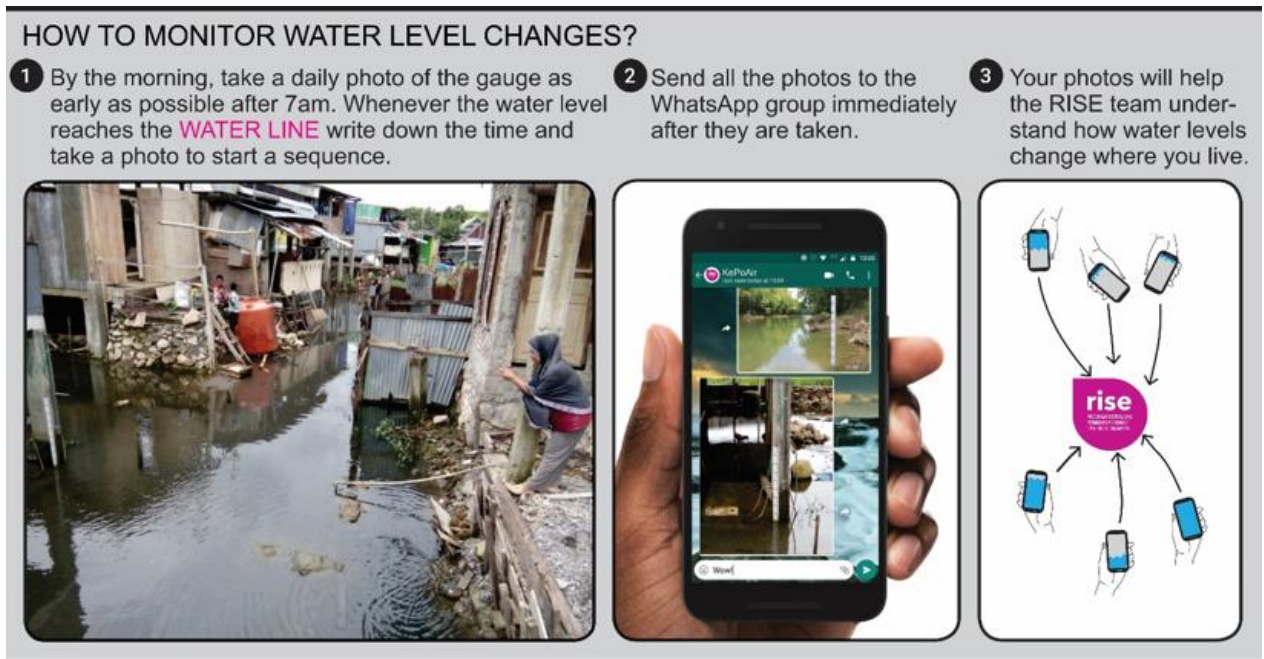

**Figure 2.** RISE's community-based flood monitoring project guidelines. These instructions were delivered and explained to participating community members in local languages to ensure data quality in the project. Source: RISE program, drawn by Daša Spasojević.

The photos were shared through a common messaging smartphone application where all volunteers were able to communicate and comment on each other's photos in a shared group. The work of monitoring water levels was done voluntarily and the only compensation offered was a monthly reimbursement to cover the access to the internet for sharing the images. Once received, the images were downloaded into a database, the water levels assessed and then recorded in a spreadsheet by the leading author as part of an action-research approach (Ramirez-Lovering et al., 2020).

Over the duration of the project, a total of 5,301 photos were received from community members in Indonesia and Fiji (n=2,433 Fiji and n=2,868 Indonesia) (Table 1). The photos allowed for comprehensive documentation of water levels in different settlements across the same catchment and, therefore, provided useful evidence of flood levels in the area (see Wolff 2021 for additional information on the data collected).

**Table 1.** Metrics of the flood-monitoring project conducted within the RISE program in Indonesia and Fiji.

| Location | Number of Settlements Monitored | Number of Participants Involved | Number of RISE Fieldworkers | Number of Gauges Installed | Total Duration of Project and Monitoring Period | Number of Photos Obtained |
|---|---|---|---|---|---|---|
| Suva, Fiji | 7 | 15 | 7 | 13 | Approx. 6 months<br>Dec 2018-Jun 2019 | 2.433 |
| Makassar, Indonesia | 6 | 11 | 4 | 7 | Approx. 8 months<br>Dec 2018-Apr 2019<br>Dec 2019-Apr 2020 | 2.868 |

The local residents who contributed to the flood-monitoring project with photos included 11 participants in Indonesia and 15 participants in Fiji (Table 1). In both countries, the groups were primarily composed of women; men only represented around a third of the collaborators, despite efforts to diversify the participants. The project was first conducted between December 2018 and the end of the wet season of 2019 and achieved significantly different outcomes in each of the countries. Following positive results of the first year in the Indonesian group, the project was repeated between December 2019 and March 2020 in Makassar.

Reflecting on these experiences, this article reflects critically on the lessons learnt during the implementation and management of the project in both countries. The methodology used to draw lessons from this project is discussed in the next section.

## 4. Method: Reflecting on the implementation of the citizen science project

Aiming to better understand the challenges and to refine the practices used in citizen science projects, this article employs an inductive approach to identify trends that emerge from the combined analysis of multiple research materials (Hodkinson, 2008). This approach is particularly suited to examine projects in which the authors are involved because it frames the investigation in a way "in which intimate knowledge and depth of understanding of the case is legitimately seen to enter the research process" (May, 2011, p. 230). As such, it is important to note the positionality of the authors as we have all been involved in different conditions with the implementation and management of the project.

Our direct involvement with the flood-monitoring project also allowed us to frame this research as a reflexive investigation (O'Reilly, 2012). Reflexive approaches are particularly important in the context of climate change studies and adaptation research as it requires researchers to be "responsive to learning and critically reflective of not only what a researcher is doing, but,[…] why, how, and to what effect" (Preston et al., 2015, p. 128). More generally, reflexive case studies that draw from context-specific knowledge are championed by researchers that argue that this kind of approach sits "at the centre of the case study as a research and teaching method; or to put it more generally, still: as a method of learning" (Flyvbjerg, 2006, p. 5). Approaching the project through a reflexive position is, therefore, a unique opportunity to draw practical lessons that can inform future similar projects. As such, the framing as a reflexive and inductive case study is not centred on the interest of providing universally generalised outcomes, but on the interest of providing lessons of particular interest to the field (Simons, 2009) derived from our own reflections and experiences.

The insights from the community fieldworkers that underpin this article build primarily on our notes from the project and a series of discussions among ourselves and between ourselves and other RISE fieldworkers. The RISE staff members involved in the project included a group of seven RISE community fieldworkers in Fiji and four in Indonesia (Table 1), as well as four researchers based in Australia directly involved with the management and analysis of the photos. These discussions were initiated as some of us collaboratively reflected on our experiences of monitoring floods within the RISE program. Building upon this initial discussion, a semi-structured questionnaire (Gilbert, 2008) was created to further explore the different experiences in both countries. This questionnaire consisted of eight open-ended questions to encourage the members of the teams in Fiji, Indonesia and Australia to reflect on the stages of implementation, engagement and analysis of the results of the project.

The answers to the questions were analysed to identify the differences between the Indonesian and Fijian experiences with the project. They were analysed separately to identify common themes and later compared to draw on the lessons pointed out by the teams in each country. These experiences were then compared to the results of the project in terms of the total number of photos obtained and community engagement. Since the participants were asked to contribute with photos every day, we considered that the frequency of photos is the main indication of community engagement in the flood-monitoring project. To ensure that the results of this analysis were representative they were synthesised in the next section and reviewed by members of RISE fieldworkers' teams in both countries, who co-authored this article.

The experiences of RISE fieldworkers, who were in direct contact with the participants throughout the project, were particularly important to identify the main challenges and insights from the project. The fieldworkers are all citizens of the countries in which they work and are familiar with the language and cultural context of the communities. They come from a range of disciplinary backgrounds, including architecture, engineering and community development. Beyond other responsibilities within RISE, they were involved in the flood monitoring campaign through activities such as flood gauge installation and community engagement (training and managing daily reporting). Based on their experiences, we discuss how effective the initiative was at engaging community members and collecting frequent flood data. The names of participants and fieldworkers involved are not disclosed in this article to protect their identities.

## 5. Results

The analysis of the answers provided by the Indonesian and Fijian fieldworker's teams revealed important differences in the approaches used in the implementation of the project in both countries. Figure 3 illustrates the duration of the flood-monitoring project in Indonesia and Fiji and highlights the days in which participants monitored each of the gauges. This analysis of the engagement in the project shows that while the flood-monitoring project in Fiji had received photos from all gauges by the end of January of 2019, the daily updates were significantly more irregular in this group throughout the whole monitoring period. The fieldworkers involved in the project in both countries shared insights into some of the practical and operational aspects behind these results which were grouped into four main findings, discussed in the following sections.

### 5.1 Pre-existing Relationships and Interests

In both countries, the communities were enthusiastic and demonstrated significant interest in the project. The Fijian fieldworkers mentioned that the project was initially well received by the communities because they were excited about the RISE program. According to them, the communities fully supported the efforts and goals of the flood-monitoring project because they were aware that the monitoring would be used to inform the design of infrastructure systems. In Indonesia, the fieldworkers shared that due to other engagement activities conducted within the broader RISE research agenda, there were pre-existing relationships between the fieldworkers and the communities. As a result, the team had been in contact with the residents for more than a year before the beginning of the flood-monitoring project which might have facilitated the communication with community members. These reflections suggest that the broader RISE program and pre-existing relationships with the communities made training and initial engagement of community members in the activity easier than expected.

Participating community members also shared with the fieldworkers that they had an interest in the flood monitoring initiative because they were already aware of the flood-prone nature of the sites. According to the fieldworkers in both countries, the interest in monitoring floods was already present in the communities as evidenced by the fact that community members were already sharing anecdotal flood descriptions and sending flood photos to RISE engagement team even before the project. According to fieldworkers in Indonesia, most of the participants understood the main purpose of registering frequent water level variations for the RISE program. As such, they were in general supportive of the activity and approached it as a platform for improving communication and sharing flood information. The fieldworkers also suggested that the severity of the floods experienced by the communities in Makassar might have resulted in a greater interest in the project in Indonesia. The flood-monitoring project, therefore, can be seen as a platform that allowed for a more systematic collection of data related to an already existing interest within the community.

### 5.2 Participant Selection

In both Suva and Makassar, the fieldworkers found that the selection of participants played an important role in the success of the project. In both countries, the selection of participants was primarily conducted by the fieldworkers with the support of local community leaders and considered, first of all, the proximity of participants to a particularly flood-prone area. In Fiji, the fieldworkers highlighted that the participant selection was challenging because they had to consider whether residents had access to a smartphone capable of joining the messaging application. Additionally, they mentioned that most community members approached were not able to participate in the flood monitoring project due to work commitments and competing priorities.

The Indonesian fieldworkers identified that the popularity of mobile internet and social media in the country were central to the success of the project. The pre-existing interest in sharing flood photos and accounts through social media meant that the flood-monitoring project became an additional platform for a practice that was already part of the local response to floods. The use of an accessible and familiar platform, which was already commonly used by the participants, also facilitated the process. These accounts suggest that the equity of access, familiarity with the platform and availability of resources were still obstacles for the project and, as a result, not all residents were equally interested or able to join the initiative.

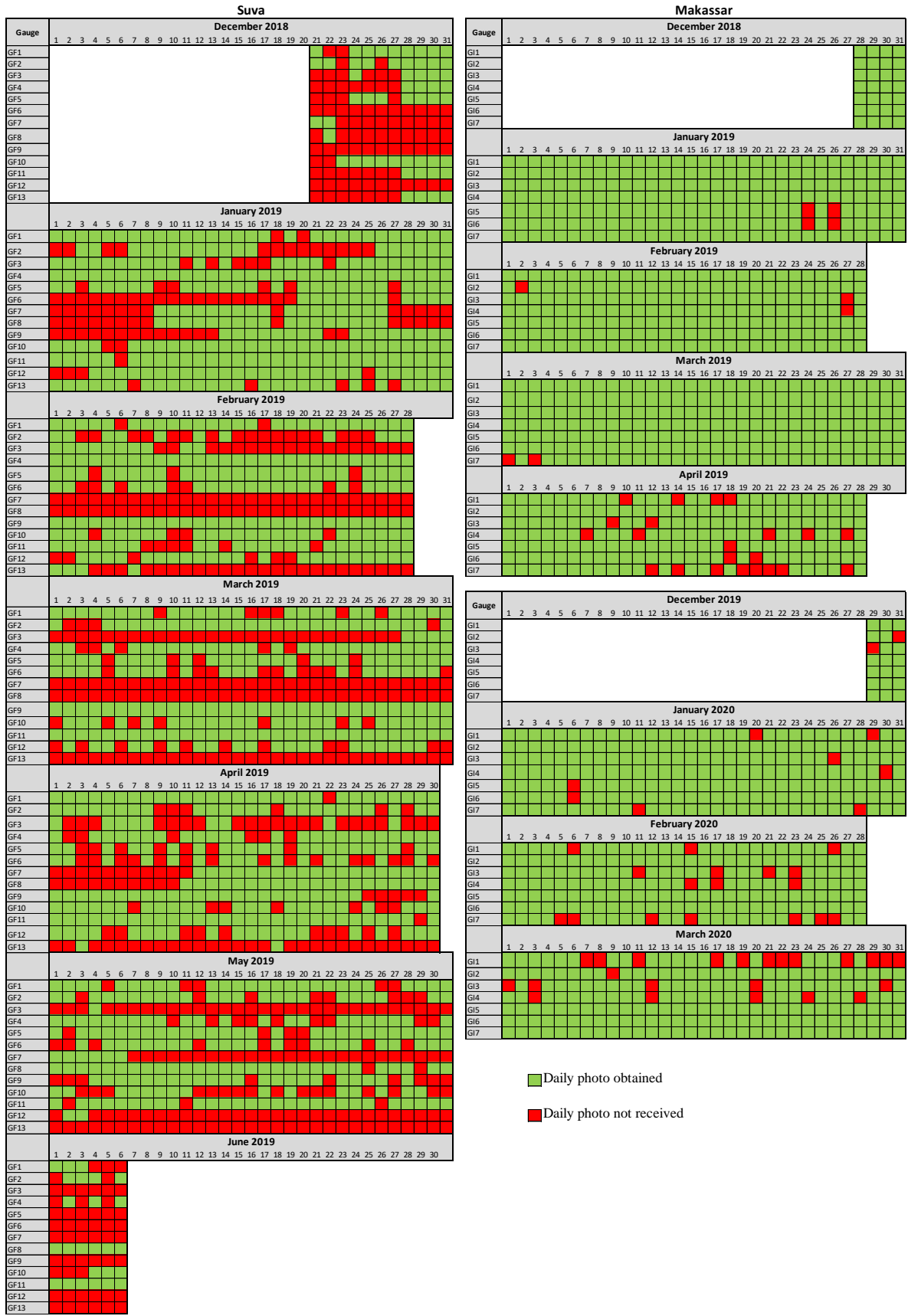

**Figure 3.** Engagement in RISE's flood-monitoring project in Suva (Fiji) and Makassar (Indonesia). The rows represent the different gauges and the columns represent the days of monitoring. The cells shaded in green illustrate days in which a photo was received for a particular gauge and are, therefore, representative of the engagement in the project.

### 5.3 Group dynamics and support by Fieldworkers

The frequency of interactions through the monitoring group was also considered an important factor influencing the engagement in the flood-monitoring project in both countries. While the Indonesian team mentioned that daily feedback and participation in the sharing platform was an important strategy to ensure positive results, the Fijian team followed up with the participants weekly when they noticed signs of disengagement. The Fijian fieldworkers shared that the engagement model could be improved in a future iteration of the project by assigning fewer fieldworkers to mediate the activity and by more closely communicating with residents. During the implementation in Fiji, each fieldworker was assigned to follow up with the participants of one settlement. This was a significant contrast in relation to how the project was conducted in Indonesia, where the communication with participants was conducted primarily by a couple of fieldworkers who were in charge of the engagement of all participants in the activity. According to the team, the involvement of several staff members in the project in Fiji meant that there were times in which certain RISE staff were not available to follow up on their communities creating a discontinuity in the monitoring. The fieldworkers in the Indonesian project, however, would follow up with community members very frequently and encourage engagement by sharing weather and flood-related information of interest in the group daily.

Additionally, other aspects not previously foreseen might have influenced the varying degrees of engagement observed in the two groups. According to the fieldworkers, the compensation for the internet usage and acknowledgement of the participant's contributions was generally enough to ensure that community members were able to participate. In Fiji, however, the fieldworkers identified that participants would sometimes be unable to contribute to the initiative if they did not have any support in their home or community to take pictures on their behalf when they were unavailable. The weekly provision of phone credit recharges (for internet access) was also considered a challenge for the fieldworkers in Fiji. The fieldworkers revealed that the recharge credits would sometimes expire before the end of the week preventing participants from sharing photos consistently. It is important to highlight that the Fijian flood-monitoring group was also considerably larger, monitoring 13 gauges, while the Indonesian was overseeing 7 gauges. These accounts reiterate the importance of considering the unequal access to technology and the accessibility of the platform used in citizen science projects.

### 5.4 Value for communities and participants

Fieldworkers in Indonesia suggested that engagement in the project played an important role in strengthening local flood response mechanisms. Following a major flood in early 2019 that was registered by the flood-monitoring project, residents were able to reach out to local support networks such as the aid from higher-income neighbours and local institutions. They described the community's interest in having access to the results, and their belief that monitoring floods would be important not only for decision-makers but also for the communities suggesting that the data-collection is not unidirectional in benefit.

The fieldworkers also shared the belief that comprehensive and structured documentation of water level fluctuations is valued by the community. Their accounts suggest that the flood monitoring project can be a tool to understand the local flood dynamics and advocate for governmental support using the collected evidence. As such, the fieldworkers communicated needs and interests between communities and RISE researchers, revealing that the project has the potential to contribute to future advocacy with decision-makers in local government. Since the citizen science project made the results accessible to the community, we believe this will serve as a household decision-making tool for future buildings and community action in the years to come. The flood documentation is now also accessible to elected community leaders and can be used as a resource to support political action.

### 6. Discussion

The findings indicate that the pre-existing relationship between the RISE Program and these communities and the selection of participants was critical for the success of the flood monitoring project. The differences observed between Fiji and Indonesia suggest that the sustainability of the project in the long term is highly dependent on access to resources and familiarity with the technology used. These findings are commensurate with the recommendations identified by other authors who explored the challenges of citizen science (Conrad & Daoust, 2008) and mapped how communities can participate in flood governance (Mees et al., 2017).

While the project was restricted to a predetermined duration in the case of RISE, the fieldworkers identified other local stakeholders that could be able to support the continuation of a future iteration of the project locally. Fieldworkers in Indonesia argued that citizen science projects would be of great interest to local authorities, following the precedent of the PetaBencana project (Fadmastuti, 2019). The BNPB (National Disaster Management Authority) and the BPBD (Regional Disaster Management Authority), in particular, are natural partners for such projects. This is aligned with the findings of other authors who highlight the need to identify local stakeholders that can further support citizen science projects in the long term (Conrad & Hilchey, 2011; Legg & Nagy, 2006).

The findings also suggest that it is essential to carefully consider how citizen science projects offer the data back to the community. In order to design the project to works as a platform for mutual knowledge transfer, the community should have access to all relevant results of the project. This is particularly relevant if the results can help residents manage disaster risk at the local scale by being incorporated within local community-based disaster risk reduction strategies (Shaw, 2016). The reflections of the fieldworkers involved in RISE flood-monitoring project indicate that further strategies should be developed to make reports more accessible, ensuring that they are available and easy to understand by the wider community. To date, the fieldworkers agreed that the main legacy of the project in the long term is the designed infrastructure located above flood levels, which is directly beneficial to the community.

Critical analysis of the project also reveals other aspects not previously foreseen that influenced the varying degrees of engagement observed in the two groups. According to the fieldworkers, most participants in Fiji reported not being familiar with the messaging application nor with the process of sharing photos through the internet before the project. The participants from Makassar, conversely, demonstrated being more comfortable and adept at using the messaging application. This situation reiterates the importance of considering the unequal access to technology and the accessibility of the platform used in citizen science projects (Assumpção et al., 2018; McCallum et al., 2016). This indicates that the process of engaging participants in citizen science must take into consideration existing communication practices and technologies that community members already use in their everyday lives.

It is also worth noting that to improve engagement and ensure that more gauges were monitored consistently, RISE's flood-monitoring framework gradually transformed to better communicate with the participants. The need to adapt frameworks and practices to local contexts is well-documented in the field of planning (Healey, 2007) and has gained traction within the field of citizen science (Cheung & Feldman, 2019; Porto de Albuquerque & Albino de Almeida, 2020). In the case of RISE, the fieldworkers noticed that Indonesian participants valued receiving monthly reports in which their contributions were acknowledged. Learning from the feedback from participants, the team recognised one of the participants as the "contributor of the month" who was acknowledged publicly. This was identified as one of the reasons why the project managed to effective gather flood data in Indonesia.

**7. Conclusions**

This article provided insights into how projects can engage communities in the management of floods based on the experiences of a citizen science project within the RISE program. It did so by documenting the process of implementation of the project and reflecting on the experiences of the fieldworkers that involved communities in the monitoring of floods in Indonesia and Fiji between 2018 and 2020. Our findings contribute to the growing body of literature regarding the potentials of citizen science as a valuable tool to promote local action and local knowledge creation.

This case study suggests important operational aspects to fulfil the potential of participatory flood monitoring and mapping practices to make room for vulnerable communities to have an active voice in city planning (Miraftab, 2009). According to the fieldworkers, the approach significantly expanded local knowledge of environmental threats and provided the community with a structured flood record, which could be continued independent of RISE using the gauges installed by the project.

The analysis of the engagement in the sharing platform was also useful to reveal how effective the project was in creating a collaborative and purposeful platform. For instance, the frequent exchange of messages between participants of the flood-monitoring group in Makassar before and during floods suggests that the platform performed other roles other than serving as a repository of photos. This situation is evidenced by the use of the citizen science group as a sharing platform through which participants warned others about the weather forecasts, discussed news and exchanged information relevant to the surrounding settlements.

The findings suggest that citizen science can support data collection, but it requires resources, technical expertise, and mediation that might not be fully available in the most disadvantaged contexts. As such, the study highlights that citizen science can support data collection, community engagement and risk awareness (Cheung & Feldman, 2019; Marchezini et al., 2017), but it should not be seen as a "solution" to the systemic and structural issues that underpin existing vulnerabilities. Consequently, the benefits of community engagement in flood data collection should not be seen as an opportunity to transfer responsibilities for flood management from governments to already vulnerable and historically disadvantaged communities.

These conclusions are particularly relevant for other projects that propose the use of citizen science for leaving "no one behind" in the context of communities living in vulnerable conditions. The experiences of the flood-monitoring project within RISE suggest that citizen science projects must be oriented by an interest in inclusive planning practices that account for the unequal access to resources and expertise in particular contexts. While citizen science is not expected to resolve the systemic roots of vulnerability in informal settlements (Rocco & Van Ballegooijen, 2018), it can contribute to addressing data gaps that are expected to be further aggravated by the interactions between climate change and rapid urban development. Contributing to a growing body of knowledge that argues for a "citizenship from below" (Marfai et al., 2015; Roy, 2005), citizen science can be a successful tool in the process of raising awareness and creating momentum for more inclusive practices in flood risk management.

**Acknowledgments**

The authors acknowledge the contributions of the colleagues within the RISE Consortium, particularly the researchers and staff members involved in the flood-monitoring project in Fiji and Indonesia including Isoa Vakarewa, Alex Wilson and Mosese Walesi, without which the flood-monitoring program here discussed would not have been possible. We would also like to thank the contribution of all residents that participated in this project.

The RISE program is funded by the Wellcome Trust [OPOH grant 205222/Z/16/Z], the New Zealand Ministry of Foreign Affairs and Trade, the Australian Department of Foreign Affairs and Trade, the Asian Development Bank, the Government of Fiji, the City of Makassar and Monash University, and involves partnerships and in-kind contributions from the Cooperative Research Centre for Water Sensitive Cities, Fiji National University, Hasanuddin University, Southeast Water, Melbourne Water, Live and Learn Environmental Education, UN-Habitat, UNU-IIGH, WaterAid International and Oxfam.

Ethics review and approval was secured by the Monash University Human Research Ethics Committee (Melbourne, Australia), and the Ministry of Research, Technology and Higher Education Ethics Committee of Medical Research at Universitas Hasanuddin (Makassar, South Sulawesi, Indonesia); and Fiji National University Ethics Committee. The RISE trial is registered on the Australian New Zealand Clinical Trials Registry (ANZCTR) (Trial ID: ACTRN12618000633280). All study settlements, households, and caregivers/respondents provided informed consent.

**Conflicts of Interests**

The authors declare no conflict of interest.

![Cogitatio]

**About the Authors**

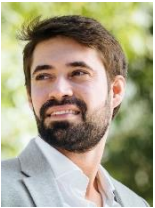

Erich Wolff is an interdisciplinary researcher and educator with a civil engineering and architecture background. His research delves into the concepts of risk and disaster to interrogate how these frameworks inform the production of infrastructure and the built environment. His recent work on citizen science investigates the role of community-based methods and the relationship between researchers and communities in the process of making sense of floods in historically disadvantaged contexts.

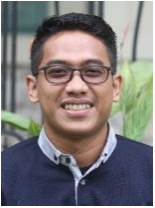

Noor Ilhamsyah graduated from Hasanuddin University, pursued a research way in the architecture degree and has worked as an architect and design consultant in several firms and projects in Sulawesi, Indonesia. Currently works as an architect who engages communities to be actively involved in the design and construction processes within RISE. In citizen science he played an active role to set up the instruments, involving the community and facilitating the flood monitoring process.

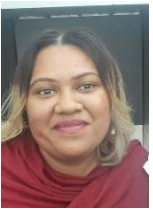

Mere Jane Sawailau is a Community Fieldworker within the RISE Program in Fiji. She has a bachelor degree in Sociology and Social Work and is a Gold Medalist in Social Work. She is passionate about community development work, advocacy and working with vulnerable people in society. Her goal is to be the best version of herself in community work and to further her studies and hope to inspire other iTaukei women in her community. She aspires to help vulnerable people to empower themselves through the work that she does.

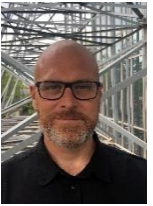

Diego Ramírez-Lovering is Professor of Architecture and Director of the Informal Cities Lab (ICL) at Monash University. The Lab undertakes design-based research exploring and speculating on the conditions of informality in developing cities in the Global South and through a lens of planetary health. ICL research — designed and conducted in collaboration with government and industry — strives for impact. It purposefully and strategically targets implementation at the intersection of academic research and international development with a key focus on the Indo-Pacific region.

Dr Matthew French is a development practitioner specialising in affordable housing, slum upgrading and community planning. Matthew's research and practice explores the interrelationship between environmental and socio-cultural dimensions of sustainable urban development. He has designed and implemented urban development programs in Africa, Latin America, the Middle East and the Asia-Pacific.